# Network tomography based on 1-D projections

**Aiyou Chen[1] and Jin Cao[1]**

*Bell Laboratories, Alcatel-Lucent Technologies*

**Abstract:** Network tomography has been regarded as one of the most promising methodologies for performance evaluation and diagnosis of the massive and decentralized Internet. This paper proposes a new estimation approach for solving a class of inverse problems in network tomography, based on marginal distributions of a sequence of one-dimensional linear projections of the observed data. We give a general identifiability result for the proposed method and study the design issue of these one dimensional projections in terms of statistical efficiency. We show that for a simple Gaussian tomography model, there is an optimal set of one-dimensional projections such that the estimator obtained from these projections is asymptotically as efficient as the maximum likelihood estimator based on the joint distribution of the observed data. For practical applications, we carry out simulation studies of the proposed method for two instances of network tomography. The first is for traffic demand tomography using a Gaussian Origin-Destination traffic model with a power relation between its mean and variance, and the second is for network delay tomography where the link delays are to be estimated from the end-to-end path delays. We compare estimators obtained from our method and that obtained from using the joint distribution and other lower dimensional projections, and show that in both cases, the proposed method yields satisfactory results.

## 1. Introduction

Network performance monitoring and diagnosis is a challenging problem due to the size and the decentralized nature of the Internet. For instance, when an end-to-end measurement indicates the performance degradation of an Internet path, the exact cause is hard to uncover because the path may traverse several autonomous systems (AS) that are often owned by different entities, e.g., a service provider, and the service providers generally do not share their internal performance. Even they do, there is no scalable way to correlate the link level measurements to end-to-end performance in a large network like the Internet. Similarly, the service providers may be interested in the end-to-end path characteristics that they can not observe directly.

Network tomography is a technology for addressing these issues [1, 4, 5, 21] (see [8] for an excellent review of this topic). The key advantage of network tomography is that it does not require the collaboration between the network elements and the end users. There are two main classes of problems being studied in the literature. The first estimates the link-level characteristics, such as packet loss or delay based on end-to-end measurements [1, 3, 4, 10, 13, 16, 18]. The loss tomography problem can be viewed as a special case of the delay tomography problem if loss is treated as a very large delay. Here we consider the case where packets are transmitted using the multicast protocol, that is, a packet is sent from a source to multiple destinations simultaneously. The second is traffic demand tomography where one predicts end-to-end path-level traffic intensity based on link-level traffic measurements [5, 6, 21]. For both network delay tomography and traffic demand tomography, the statistical models can be unified by

$$\mathbf{Y} = A\mathbf{X}, \tag{1}$$

where $\mathbf{X} = (X_1, \ldots, X_I)^T$ is an $I$-dimensional vector of unobservable network dynamic parameters with mutually independent components, and $\mathbf{Y} = (Y_1, \ldots, Y_J)^T$ is a $J$-dimensional vector of measurements and $A$ is a $J \times I$ routing matrix with elements 0 or 1. The objective is to estimate the distribution of $\mathbf{X}$ given independent observations from the distribution of $\mathbf{Y}$.

[1]Communications and Statistical Sciences Department, Bell Laboratories, Alcatel-Lucent Technologies, 600 Mountain Ave, Murray Hill, NJ, USA 07974, e-mail: aychen@research.bell-labs.com; cao@research.bell-labs.com

As with other inverse problems, the main difficulty in network tomography lies in the fact that $I > J$. For the traffic demand tomography $I$ can be as large as $J^2$, and for the delay tomography model, $I$ can be as large as $2J - 1$. As a result, $A$ is not invertible and the tomography problem is ill-posed. However, if components of $\mathbf{X}$ are assumed independent, it can be shown that the distribution functions of $\mathbf{X}$ under model (1) can be uniquely determined up to their means under mild conditions [9]. This mean ambiguity can be further removed if the component distributions of $\mathbf{X}$ satisfy some additional constraints, for example, a relation between their means and higher order moments (such as the Poisson distribution), positive point mass at zero etc [5, 12, 19, 21].

To estimate the distribution of $\mathbf{X}$, likelihood based inference has been proposed [4, 5, 15, 16]. However, since the likelihood involves a high order convolution, such inference is computationally expensive except for specific distributions. This can be limiting in some circumstances. For example, in delay tomography where the link delays are to be estimated, the continuous distributions for both end-to-end and link delays are approximated by non-parametric discrete distributions using the same set of bins of equal widths. This is problematic for a heterogeneous network such as Internet where the link delay distributions can differ significantly. To overcome the computational difficulties in likelihood based inference, a characteristic-function based generalized moment estimator has been proposed for general distributions [9], where the model parameters are estimated by minimizing a contrast function between the empirical characteristic function and the parametrized characteristic function of $\mathbf{Y}$ under the model.

However, it has been realized that either the full likelihood or joint characteristic function based inference may still be expensive when the dimension of $\mathbf{X}$ is high. A solution to this is the *pseudo-likelihood* approach proposed in [16], where the parameters are estimated by minimizing a pseudo-likelihood function that is constructed by multiplying the marginal likelihoods of a sequence of subsets $\mathbf{Y}^s$ of the high-dimensional observation $\mathbf{Y}$. The idea is that these marginal likelihoods only focus on a small subset $\mathbf{Y}^s$ and thus are computationally much cheaper. Specifically, the authors considered constructing these subsets using all pairwise components of $\mathbf{Y}$, i.e., $\mathbf{Y}^s = (Y_{j_1}, Y_{j_2}), 1 \le j_1 < j_2 \le J$, and found that such a pseudo-likelihood based estimator is computationally efficient as compared to the estimator based on full likelihood meanwhile has little loss in statistical efficiency. The same idea has been used in the characteristic-function based estimator in [9] by considering characteristic functions of a sequence of low dimensional subsets $\mathbf{Y}_s$, and in the local likelihood estimator in [15] by considering $\mathbf{Y}_s$ of both one and two dimensions.

In all the above approaches, each $\mathbf{Y}^s$ can be considered as a projection of a high dimensional measurement $\mathbf{Y}$ onto a low dimensional subspace, taken along the axis of components of $\mathbf{Y}$. One can further generalize this and take projections in arbitrary directions, for example, using $B_s^T\mathbf{Y}$ where $B_s$ is matrix with a small number of columns. However, an optimal choice of these lower-dimensional projections so as to balance the computational complexity and statistical efficiency has not been studied previously. For statistical efficiency, one might want to use relatively high dimensional projections so that the information on multivariate dependency will not be lost. For computational efficiency, one might prefer to a small set of lower dimensional projections.

This paper provides a partial solution to the design issue of these lower dimensional projections. Here we consider the extreme case – one-dimensional (1D) linear projections of the observed data $\mathbf{Y}$. That is to say, the statistical inference regarding the distribution of $\mathbf{X}$ is based on the marginal distributions of a series of 1D-projections of $\mathbf{Y}$, say $\beta_k^T\mathbf{Y}$, $\beta_k \in \mathcal{R}^J$ for $k = 1, \ldots, K$. The contributions of this paper are two-fold. First, we give a sufficient condition for the choice of these 1D-projections so that the $\mathbf{X}$ distribution can be uniquely determined. Second, we study the design issue of such 1D-projections in terms of statistical efficiency. For a simple Gaussian tomography model where component distributions of $X$ are Gaussian, we show that there exists an optimal choice of 1D-projections, selected by a correlation based rule, from which the obtained estimator is asymptotically as efficient as the maximum likelihood estimator (MLE) based on the joint distributions of $\mathbf{Y}$. For more realistic tomography models, we carry out simulation studies of two instances: the first is for traffic demand tomography where the Origin-Destination traffic is also Gaussian but its mean and variance are related through a power equation, and the second is for network delay tomography where the link delays are to be estimated using a continuous mixture distribution. For both cases, we show the proposed method based on 1D-projections yields satisfactory results as compared to estimators using other choice of projections and the complete data.

The remaining of the paper is organized as follows. In Section 2, the method of 1D-projections is proposed, and the identifiability issue and parameter estimation are discussed. In Section 3, a simple Gaussian tomography model is analyzed, and the optimal design of 1D-projections and its efficiency are studied. In Section 4, simulation studies of the 1D-projection method are presented for traffic demand tomography and network delay tomography. We conclude in Section 5.

The following conventions will be used throughout the paper. 1D-projectio- ns represent one-dimensional projections with the form $\beta_k^T\mathbf{Y}$. 2D-projections represent pairwise components of $\mathbf{Y}$, e.g. $(Y_j, Y_{j'})$. A lower case letter represents a vector and an upper case letter represents a matrix, with the exception of $\mathbf{X}$ and $\mathbf{Y}$, which represent random vectors as in model (1). $M_{ab}$ or $M(a, b)$ represents the $(a, b)$th element of a matrix $M$. $v_i$ is the $i$th element of a vector $v$ and $\beta_k \in \mathcal{R}^J$ is a column vector and $\beta_{ki}$ is its $i$th element. $M^T$ and $v^T$ represent the transpose of $M$ and $v$ respectively. $M^{-T}$ is the transpose of $M^{-1}$, the inverse of $M$.

## 2. Method of one-dimensional projections

In this section, we formally describe the method of 1D-projections for solving the inverse problem in (1) in the context of network tomography. We first present a necessary and sufficient condition for identifiability and then discuss the parameter estimation in this setting.

### 2.1. Identifiability

One fundamental question of the 1D-projection method is whether the distribution of $\mathbf{X}$ can be uniquely determined by the marginal distributions of these 1D-projections. This is the so-called *identifiability* issue. For simplicity we shall start with a simple matrix $A$ derived from the two-leaf tree delay tomography model (Figure 1), and then generalize it to an arbitrary matrix. For illustration purpose, we use a special set of 1D-projections on the two-leaf tree to explain the main idea behind the identifiability. For the two-leaf tree in Figure 1, let $\mathbf{X} = (X_1, X_2, X_3)^T$ be the internal link delays from top to bottom and left to right, and $\mathbf{Y} = (Y_1, Y_2)^T$ be the end-to-end delay from the root node to the two leaves from left to right. Since the end-to-end delay is the sum of internal link delays on the path, we can write $Y_1 = X_1 + X_2$ and $Y_2 = X_1 + X_3$. That is, $A = [1, 1, 0; 1, 0, 1]$. Following [9], we assume that the characteristic function of each component of $\mathbf{X}$ is analytic[1] and we refer to this as the *analytic condition* in this paper.

**Lemma 2.1.** *For the two-leaf tree in Figure 1, assume that* $\mathbf{X}$ *has mutually independent components and satisfies the analytic condition. Then the distribution of* $\mathbf{X}$ *is determined up to a shift in its mean by the marginal distributions of* $Y_1, Y_2, Y_1 + aY_2$ *if* $a \neq 0$ *and* $a \neq -1$, *where the mean of* $\mathbf{X}$ *satisfies the constraint* $E[X_1] + E[X_2] = E[Y_1]$ *and* $E[X_1] + E[X_3] = E[Y_2]$.

*Proof.* Let $\beta_1 = (1, 0)^T$, $\beta_2 = (0, 1)^T$ and $\beta_3 = (1, a)^T$, then the three projections can be written as $\beta_k^T \mathbf{Y}$, $k = 1, 2, 3$. Let $\gamma_k = A^T \beta_k = (\gamma_{k1}, \gamma_{k2}, \gamma_{k3})^T$, then the characteristic function of $\beta_k^T Y$ is equal to

$$Ee^{it\beta_k^T \mathbf{Y}} = \prod_{i=1}^{3} \psi_i(\gamma_{ki} t),$$

where $\psi_i$ is the characteristic function of $X_i$. Suppose that there exists another set of characteristic functions $\bar{\psi}_i$ which also satisfy the above three equations. Let $\omega_i(t) = \log \psi_i(t) - \log \bar{\psi}_i(t)$. Then we have for $k = 1, 2, 3$,

$$\sum_{i=1}^{3} \omega_i(\gamma_{ki} t) = 0.$$

Let $d_{in}$ be the $n$th order derivative of $\omega_i(t)$ at $t = 0$. By evaluating the $n$th order derivatives at zero, we have for $k = 1, 2, 3$,

$$\sum_{i=1}^{3} \gamma_{ki}^n d_{in} = 0.$$

For each $n \geq 2$, let $M_n$ be the coefficient matrix of the above linear equations of $\{d_{in} : i = 1, 2, 3\}$. Since $A = [1, 1, 0; 1, 0, 1]$, we have

$$M_n = \begin{bmatrix} 1 & 1 & 0 \\ 1 & 0 & 1 \\ (1+a)^n & 1 & a^n \end{bmatrix}.$$

[1] An analytic characteristic function corresponds to a distribution function which has moments $m_k$ of all orders $k$ and $\limsup_{k\to\infty} [|m_k|/k!]^{1/k}$ is finite.

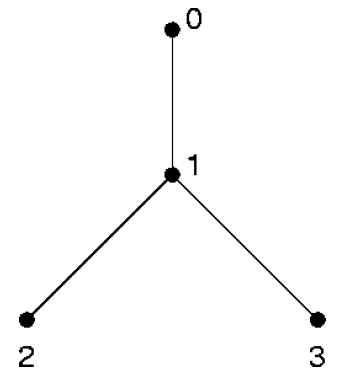

FIG 1. *Two-leaf tree*

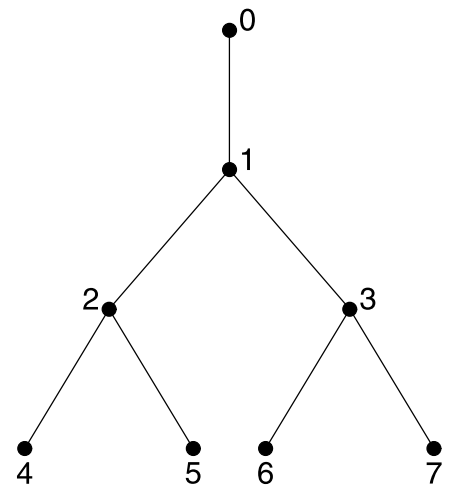

FIG 2. *Four-leaf tree*

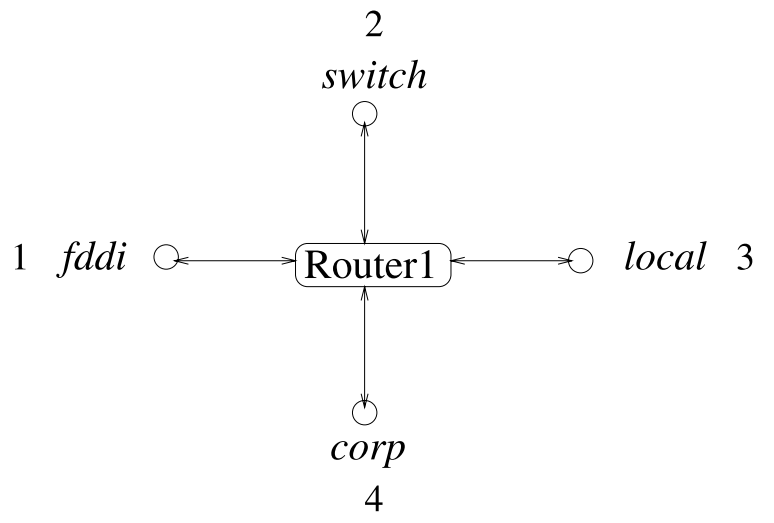

FIG 3. *One router network*

The identifiability problem is equivalent to asking whether $d_{in} = 0$ $(i = 1, 2, 3)$, i.e. $\det(M_n) \neq 0$, for all $n \geq 2$. Note that $\det(M_n) = (1 + a)^n - a^n - 1$. Let $f(x) = (1 + x)^n - x^n - 1$. Thus $f'(x) = n[(1 + x)^{n-1} - x^{n-1}]$.

If $n \geq 2$ is even, $f'(x) > 0$ and thus $f(x)$ is monotone with a unique zero $x = 0$.

If $n > 2$ is odd, $f'(x)$ has a unique zero $x = -1/2$ and thus $f(x)$ has two zeros $x = 0$ and $x = -1$.

Under assumptions of $a \neq 0$ and $a \neq -1$, $\det(M_n) = f(a) \neq 0$, and thus $d_{in} = 0$ $(i = 1, 2, 3)$ for all $n \geq 2$. Hence each $\omega_i(t)$ is a linear function. The conclusion follows readily. □

Lemma 2.1 states that if chosen properly, only three 1D-projections are needed to determine the distribution of $\mathbf{X}$ (ignoring its mean). The condition $a \neq 0$ merely says that the third projection $Y_1 + aY_2$ does not coincide with either $Y_1$ or $Y_2$. Obviously, from the proof we can see that the condition $a \neq -1$ is not needed in order to identify the even order cumulants of each $X_i$, i.e. the even order derivatives of $\log \psi_i(t)$ at $t = 0$, but $a \neq -1$ is required in order to identify the odd order cumulants.

**Remark.** It is not hard to show that all even order cumulants can be determined by the marginal distributions of arbitrary three distinct[2] projections $a_j Y_1 + b_j Y_2$, $j = 1, 2, 3$. But some further constraints, i.e. $a_j + b_j \neq 0$ $(j = 1, 2, 3)$ which is similar to the condition $a \neq -1$ in the above lemma, are needed in order to determine the odd order cumulants except the first order. Lemma 2.1 provides a simple set of 1D-projections which can be used to identify the $\mathbf{X}$ distributions. This can be easily extended to the case where $A$ corresponds to a tree topology. For a general matrix $A$, the identifiability issue is addressed by the following theorem.

**Theorem 2.1.** *For a general tomography problem $\mathbf{Y} = A\mathbf{X}$ introduced by (1), let $\beta_k^T \mathbf{Y}$, $k = 1, \ldots, K$, be $K$ 1D-projections of $\mathbf{Y}$ onto the linear space of its components. Let $M$ be a $K \times I$ matrix whose rows consist of $\beta_k^T A$, and let $M_n$ be a matrix whose elements are the nth power of the corresponding elements of $M$. Then the distribution of $\mathbf{X}$ can be determined up to the ambiguity of its mean by the marginal distributions of $\beta_k^T \mathbf{Y}$, $k = 1, \ldots, K$, if and only if $M_n$ has full column rank for all $n \geq 2$, where the mean of $\mathbf{X}$ satisfies the constraint $AE[\mathbf{X}] = E[\mathbf{Y}]$.*

The proof of Theorem 2.1 follows the same idea as that of Lemma 2.1. The details are omitted to avoid technical redundancy.

**Remark.** Theorem 2.1 provides a sufficient and necessary condition for the identifiability of the $\mathbf{X}$ distribution by using the marginal distributions of a set of

[2]Two projections are the same if one is a scale multiplication of the other.

1D-projections of the observed measurements $\mathbf{Y}$. Unfortunately, the full column rank condition on $M_n$, for any $n \geq 2$, is hard to verify for an arbitrary set of projections. Further, it is worth pointing out that for identifiability, $K \geq I$ is required since each $M_n$ has $I$ columns. However, when $A$ satisfies some sufficient conditions such that the distribution of $\mathbf{X}$ can be identified from the joint distribution of $\mathbf{Y}$ (see [9] for discussions of such $A$ matrix), then we would expect that in most cases, the marginal distributions of a set of $K = I$ properly chosen 1D-projections of $\mathbf{Y}$ can be used to identify the distribution of $\mathbf{X}$. This is due to the fact that by solving the polynomial equations, i.e. $\det(M_n) = 0$, the set of projection directions $\{\beta_k : k = 1, \ldots, I\}$ (ignoring the scales) which violates the full rank condition has Lebesgue measure zero.

### *2.2. Parameter estimation*

Let $f(\mathbf{X}|\theta)$ be the distribution of $\mathbf{X}$ with unknown parameter $\theta$. We consider two estimators of $\theta$ derived from the marginal distributions of 1D-projections $\{\beta_k^T \mathbf{Y}, k = 1, \ldots, K\}$, one based on their marginal likelihoods and the other based on their marginal characteristic functions. In later sections we shall give examples for each of these for instances of network tomography problems.

#### *2.2.1. Likelihood based inference*

Provided that it is easy to evaluate the univariate distribution of $\beta_k^T \mathbf{Y}$, the Kullback-Leibler divergence between the empirical and model distributions of each $\beta_k^T \mathbf{Y}, k = 1, \ldots, K$ can be used to obtain an estimator of $\theta$. Let $P_n$ be the empirical distribution of $\mathbf{Y}$ based on $n$ i.i.d. samples and $l_k(\cdot, \theta)$ be the logarithmic likelihood function of $\beta_k^T Y$. Similar to the maximum pseudo likelihood method in [16], an estimator of $\theta$ based on the 1D-projections can be defined by

$$\hat{\theta}_{1D} = \arg\min \sum_{k=1}^{K} \int -l_k(\beta_k^T Y, \theta) dP_n. \tag{2}$$

Usually there is no closed form solution to $\hat{\theta}_{1D}$, and the pseudo EM algorithm in [16] or other numerical optimization algorithms may be used.

#### *2.2.2. Characteristic function based inference*

Often the distribution of $\beta_k^T \mathbf{Y}$ is hard to evaluate since it is a high order convolution of the distributions of $(X_1, \ldots, X_I)$. In this case, as proposed in [9], the generalized method of moments (GMM) based on characteristic function [7, 14] can be used to obtain an estimator of $\theta$. Let $\psi_i(\cdot, \theta)$ be the characteristic function of $X_i$ and $\hat{\varphi}_k(\cdot, \theta)$ be the empirical characteristic function of $\beta_k^T \mathbf{Y}$. Let $A^i$ be the $i$th column of $A$. Then the characteristic function of $\beta_k^T \mathbf{Y}$, say $\varphi_k(\cdot, \theta)$, is equal to

$$\varphi_k(t, \theta) = E e^{it\beta_k^T \mathbf{Y}} = \prod_{i=1}^{I} \psi_i(\beta_k^T A^i t, \theta).$$

Now an estimator of $\theta$ based on characteristic functions of the 1D-projections can be defined as follows:

$$\hat{\theta}_{CF} = \arg\min \sum_{k=1}^{K} \int |\hat{\varphi}_k(t) - \varphi_k(t, \theta)|^2 d\mu(t), \tag{3}$$

where $\mu(\cdot)$ is a predetermined distribution measure on $\mathcal{R}$. In general, (3) may be solved numerically. For computational convenience, we may use an empirical distribution of $\mu$ in the above as an approximation. When each $X_i$ follows a mixture distribution, i.e. $\psi_i(t) = \sum_k p_k^i \Psi_k^i(t)$, $p_k^i \geq 0$, $\sum_k p_k^i = 1$, where $\Psi_k^i$ are characteristic functions of corresponding mixture components, and $\theta$ consists of the coefficient parameters $p_k^i$, then for each $i$, the above objective function is a quadratic function of $\{p_k^i\}$. Thus the optimization can be done iteratively by quadratic programming (see [9] for details). Asymptotic properties such as consistency and asymptotic normality have been studied for this estimator under mild conditions [7]. An improved estimator that takes into account the correlation of the empirical characteristic function $\hat{\varphi}_k$ at different points $t$ can also be considered following similar techniques developed in [9].

## 3. Optimal design of 1D projections for the Gaussian tomography model

In Section 2, it has been shown that the distribution of $\mathbf{X}$ can be identified using the marginal distributions of a set of $K$ $(K \geq I)$ properly chosen 1D-projections. In this section, we consider the optimal design of these projections. We consider the two main factors: 1) the statistical efficiency of the estimators based on these projections and 2) the computational complexity determined primarily by the number of such 1D-projections, i.e. $K$. To achieve optimal design, we first consider a simple Gaussian tomography model defined below and investigate the design issue in depth. Surprisingly, we found that for this simple model there is a minimal set of $I$ 1D-projections such that the estimator based on these projections is asymptotically as efficient as MLE. Such a set of projections constitute the optimal design.

### *3.1. The Gaussian tomography model*

The Gaussian tomography model is defined by the tomography model in (1) when the components of $\mathbf{X}$ have mutually independent Gaussian distributions, i.e., $X_i \sim \mathcal{N}(\mu_i, \theta_i)$, $i = 1, \ldots, I$, and $\mathcal{N}(\mu_i, \theta_i)$ stands for the normal distribution with mean $\mu_i$ and variance $\theta_i$. Notice that for a set of 1D-projections of $\mathbf{Y}$ that satisfies the condition of Theorem 2.1, $\theta = (\theta_1, \ldots, \theta_I)^T$ is identifiable but $(\mu_1, \ldots, \mu_I)^T$ is not because of the mean ambiguity. For simplicity assume that $\mu_i = 0$ for $i = 1, \ldots, I$. The Gaussian tomography model is defined as

$$\mathbf{Y} \sim \mathcal{N}(0, A\Theta A^T), \qquad (4)$$

where $\theta$ is the parameter of interest and $\Theta$ is a diagonal matrix with diagonal elements $\theta_i$. Since the distribution of $\beta_k^T \mathbf{Y}$ is also Gaussian and thus easy to evaluate, given a set of 1D-projections, we use the likelihood based method given by (2) in Section 2 to estimate $\theta$. We investigate its statistical efficiency as compared to that of MLE, denoted as $\hat{\theta}_{MLE}$.

Let $\Sigma = A\Theta A^T$, and let $\to_d$ indicate convergence in distribution. The following lemmas state the asymptotic distributions of the estimators $\hat{\theta}_{MLE}$ and $\hat{\theta}_{1D}$.

**Lemma 3.1.** *Suppose that $\theta$ is identifiable. Then*

$$\sqrt{n}(\hat{\theta}_{MLE} - \theta) \to_d \mathcal{N}(0, \mathcal{I}_F^{-1}),$$

*where $\mathcal{I}_F$ is the Fisher information matrix with elements $\mathcal{I}_F(a,b) = \frac{1}{2}U_{ab}^2$ and $U_{ab}$ is the $(a,b)$th element of $U = A^T\Sigma^{-1}A$.*

*Proof.* The log-likelihood function of $\mathbf{Y}$ can be written as

$$\log p_{\mathbf{Y}}(y;\theta) = -\frac{1}{2}\mathbf{Y}^T\Sigma^{-1}\mathbf{Y} - \frac{1}{2}\log(2\pi\det(\Sigma)),$$

where $\Sigma = A\Theta A^T$. Notice that $\Sigma = \sum_{j=1}^m \theta_j A^j A^{jT}$, by taking partial derivatives w.r.t. each $\theta_i$, we have

$$\frac{\partial \log p_{\mathbf{Y}}(y;\theta)}{\partial\theta_i} = \frac{1}{2}\left\{\mathbf{Y}^T\Sigma^{-1}A^iA^{iT}\Sigma^{-1}\mathbf{Y} - vec(\Sigma^{-1})^T vec(A^iA^{iT})\right\},$$

where $vec(M)$ vectorizes $M$. So the score functions, say $s_i, i = 1,\dots,I$, are

$$s_i(\mathbf{Y}) \equiv -\frac{\partial}{\partial\theta_i}\log p(\mathbf{Y};\theta) = \frac{1}{2}\left\{-(\mathbf{Y}^T\Sigma^{-1}A^i)^2 + trace(\Sigma^{-1}(A^iA^{iT}))\right\},$$

where $trace(M)$ denotes the trace of $M$. Hence the result follows from

$$\begin{aligned} cov(s_j(\mathbf{Y}), s_k(\mathbf{Y})) &= \frac{1}{4}cov((\mathbf{Y}^T\Sigma^{-1}A^j)^2, (\mathbf{Y}^T\Sigma^{-1}A^k)^2) \\ &= \frac{1}{2}\{cov(\mathbf{Y}^T\Sigma^{-1}A^j, \mathbf{Y}^T\Sigma^{-1}A^k)\}^2 \\ &= \frac{1}{2}(A^{jT}\Sigma^{-1}A^k)^2. \end{aligned}$$

The second equality in the above uses the fact that for any bivariate Gaussian random vector $(N_1, N_2)$, $cov(N_1^2, N_2^2) = 2[cov(N_1, N_2)]^2$. □

**Lemma 3.2.** *For a set of 1D-projections* $\{\beta_k^T\mathbf{Y}, k = 1,\dots,K\}$, *let* $\gamma_k = A^T\beta_k$, *and* $\sigma_k^2 = \beta_k^T\Sigma\beta_k$. *Suppose that* $\theta$ *can be identified by these 1D-projections. Then the likelihood based estimator* $\hat\theta_{1D}$ *defined by (2) using these projections has an asymptotic distribution*

$$\sqrt{n}(\hat\theta_{1D} - \theta) \to_d \mathcal{N}(0, \mathcal{C}_{1D}^{-1}\mathcal{I}_{1D}\mathcal{C}_{1D}^{-1}) \tag{5}$$

*where* $\mathcal{C}_{1D} = \frac{1}{2}V^TV$ *and* $\mathcal{I}_{1D} = \frac{1}{2}V^TWV$. *Here* $V$ *is a* $K\times I$ *matrix with elements* $V_{ka} = \sigma_k^{-2}\gamma_{ka}^2$ *and* $W$ *is a* $K\times K$ *symmetric matrix with elements* $W_{kk'} = \sigma_k^{-2}\sigma_{k'}^{-2}(\beta_k^T\Sigma\beta_{k'})^2$. *Furthermore, if* $K = I$ *and* $V$ *is invertible, then the limit covariance matrix in (5) simplifies to* $2V^{-1}WV^{-T}$.

*Proof.* Notice that

$$\beta_k^T\mathbf{Y} \sim \mathcal{N}(0, \sigma_k^2).$$

Let $l_k(\cdot,\theta)$ be the marginal logarithmic likelihood function of $\beta_k^T\mathbf{Y}$, that is,

$$l_k(z,\theta) = -\frac{1}{2}\sigma_k^{-2}z^2 - \frac{1}{2}\log(2\pi\sigma_k^2).$$

By the classical theory on M-estimators, we have

$$\sqrt{n}(\hat\theta_{1D} - \theta) \to_d \mathcal{N}(0, \mathcal{C}_{1D}^{-1}\mathcal{I}_{1D}\mathcal{C}_{1D}^{-1}) \tag{6}$$

where

$$\begin{aligned} \mathcal{C}_{1D} &= -\sum_{k=1}^K E\left[\frac{\partial^2 l_k(\beta_k^T\mathbf{Y},\theta)}{\partial\theta\partial\theta^T}\right] \\ &= \sum_{k=1}^K E\left[\left(\frac{\partial l_k(\beta_k^T\mathbf{Y},\theta)}{\partial\theta}\right)\left(\frac{\partial l_k(\beta_k^T\mathbf{Y},\theta)}{\partial\theta}\right)^T\right] \end{aligned} \tag{7}$$

and

$$(8) \qquad \mathcal{I}_{1D} = var\left[\sum_{k=1}^{K} \frac{\partial l_k(\beta_k^T \mathbf{Y}, \theta)}{\partial \theta}\right].$$

It is not hard to verify that

$$\frac{\partial l_k(z,\theta)}{\partial \theta_j} = \frac{\partial l_k(z,\theta)}{\partial \sigma_k^2}\frac{\partial \sigma_k^2}{\partial \theta_j} = \frac{\gamma_{kj}^2(z^2 - \sigma_k^2)}{2\sigma_k^4}.$$

Thus

$$\mathcal{C}_{1D}(a,b) = \sum_{k=1}^{K} \frac{1}{2}\sigma_k^{-4}\gamma_{ka}^2\gamma_{kb}^2$$

and

$$\begin{aligned}\mathcal{I}_{1D}(a,b) &= \sum_{k,k'=1}^{K} \frac{\gamma_{ka}^2\gamma_{k'b}^2}{4\sigma_k^4\sigma_{k'}^4} cov((\beta_k^T\mathbf{Y})^2, (\beta_{k'}^T\mathbf{Y})^2) \\ &= \frac{1}{2}\sum_{k,k'=1}^{K} \frac{\gamma_{ka}^2\gamma_{k'b}^2}{\sigma_k^4\sigma_{k'}^4}(\beta_k^T\Sigma\beta_{k'})^2.\end{aligned}$$

Hence, $\mathcal{C}_{1D} = \frac{1}{2}V^TV$ and $\mathcal{I}_{1D} = \frac{1}{2}V^TWV$. □

### *3.2. Optimal projections*

Given the asymptotic variance $\mathcal{C}_{1D}^{-1}\mathcal{I}_{1D}\mathcal{C}_{1D}^{-1}$ of the estimator $\hat{\theta}_{1D}$ (6, 7 and 8), it is not obvious how one can choose a set of 1D-projections so that the asymptotic covariance matrix can be minimized. By (7) $\mathcal{C}_{1D}$ is the summation of the information matrix of individual projection $\beta_k^T\mathbf{Y}$. Note that $cov((\beta_k\mathbf{Y})^2, (\beta_{k'}\mathbf{Y})^2) > 0$. By (9), this implies that the score functions of individual 1D-projections, i.e. $\frac{\partial l_k(\beta_k^T\mathbf{Y},\theta)}{\partial\theta}$ $(k = 1, \ldots, K)$, are positively correlated. Intuitively, the directions $\beta_k$ should be chosen such that the scores and thus the projections $\beta_k^T\mathbf{Y}$ are mutually uncorrelated as much as possible from each other. To be more precise, reduce $\mathcal{I}_{1D}$ significantly while keep $\mathcal{C}_{1D}$ stable. Since each projection is a linear function of the components of $\mathbf{Y}$ and the individual components of $\mathbf{X}$ are mutually independent, thus there will not be much information overlap if the projections are chosen such that each is close to an individual component of $\mathbf{X}$. This motivates the following design for the Gaussian tomography model.

Take $K = I$. For each $k \in \{1, \ldots, I\}$, pick $\beta_k$ such that the correlation between $\beta_k^T\mathbf{Y}$ and $X_k$, say $corr(\beta_k^T\mathbf{Y}, X_k)$, is maximized. Since

$$corr(X_k, \beta^T\mathbf{Y}) = \frac{cov(X_k, \beta^T A\mathbf{X})}{\sqrt{var(X_k)var(\beta^T\mathbf{Y})}} = \frac{\beta^T A^k \theta_k^{1/2}}{\sqrt{\beta^T cov(\mathbf{Y})\beta}},$$

the scale on $\beta$ does not affect the correlation. Furthermore, notice that the scale does not affect either $\mathcal{C}_{1D}$ or $\mathcal{I}_{1D}$. By assuming $var(\beta_k^T\mathbf{Y}) = 1$, $\beta_k$ can be determined :

$$(9) \qquad \beta_k = \lambda_k^{-1}\Sigma^{-1}A^k,$$

where $\lambda_k = (A^{kT}\Sigma^{-1}A^k)^{1/2}$. Theorem 3.1 below shows that the projection directions chosen by the above correlation rule leads to an efficient estimator of $\theta$.

**Theorem 3.1.** *Under the same condition as Lemma 3.1, if the 1D-projections are taken by the maximum correlation rule given by (9), then for the simple Gaussian tomography model, the 1D-projection estimator given by (2) is asymptotically as efficient as MLE.*

*Proof.* By (9), $\gamma_{ka} = \lambda_k^{-1} A^{aT}\Sigma^{-1}A^k$. Plugging this into (5), it is not hard to verify that

$$V = 2S^{-1}\mathcal{I}_F,$$

and

$$W = 2S^{-1}\mathcal{I}_F S^{-1},$$

where $\mathcal{I}_F$ is the same as in Lemma 3.1 and $S$ is a diagonal matrix with diagonals $\lambda_k^2$, $k = 1, \ldots, I$. Thus $V$ is invertible and the limit covariance matrix in (5) simplifies to $\mathcal{I}_F^{-1}$. The result follows. □

Notice that the optimal 1D-projections $\beta_k$s in (9) depend on unknown parameters $\theta$ and thus are unavailable. Fortunately, $\beta_k$ only depends on $\Sigma$ and the sample covariance of $\mathbf{Y}$, say $\hat{\Sigma}$, is a $\sqrt{n}$-consistent estimator of $\Sigma$. Thus we can plug-in $\hat{\Sigma}$ and obtain empirical estimates of $\beta_k$. By assuming that $\theta$ belongs to a compact subset of $\mathcal{R}^I$, from the theory of generalized M-estimators in [2], this plug-in 1D-projection estimator is asymptotically efficient. We omit the tedious technical verification but refer to Chapter 7 of [2] for details.

More realistic Gaussian tomography models assume a mean-variance relationship and thus the above theorem may not hold. But simulation studies in the next section suggest that the above plug-in 1D-projections still work better than arbitrary projections and the performance of the corresponding estimator is close to MLE.

### *3.3. Comparison with other projections*

We now compare the statistical efficiency of $\hat{\theta}_{1D}$ based on the optimal set of 1D-projections given by (9) with that based on other choices of projections. We consider two such choices. The first is based on the set of all pairwise 2D-projections of $\mathbf{Y}$ that are proposed in [16], called $\hat{\theta}_{2D}$, the second is based on a set of $K$ randomly chosen 1D-projections while adjusting for the correlation of $\mathbf{Y}$, i.e., each random projection is generated by

$$\beta_k^T\mathbf{Y} = \alpha_k^T\Sigma^{-1/2}\mathbf{Y}, \tag{10}$$

where $\alpha_k$ is drawn independently from the standard multivariate Gaussian distribution with an identity covariance matrix.

Let $\hat{\theta}_{2D}$ be the estimator of $\theta$ based on all pairwise 2D-projections of $\mathbf{Y}$ and let $f((Y_k, Y_{k'})|\theta)$ be the distribution of the bivariate variable $(Y_k, Y_{k'})$. Then $\hat{\theta}_{2D}$ is defined by

$$\hat{\theta}_{2D} = \arg\min_\theta \int \sum_{1\le k<k'\le J} - \log f((Y_k, Y_{k'})|\theta)dP_n. \tag{11}$$

Following similar arguments as in Lemma 3.2, it can be shown that

$$\sqrt{n}(\hat{\theta}_{2D} - \theta) \to_d \mathcal{N}\left(0, \mathcal{C}_{2D}^{-1}\mathcal{I}_{2D}\mathcal{C}_{2D}^{-1}\right), \tag{12}$$

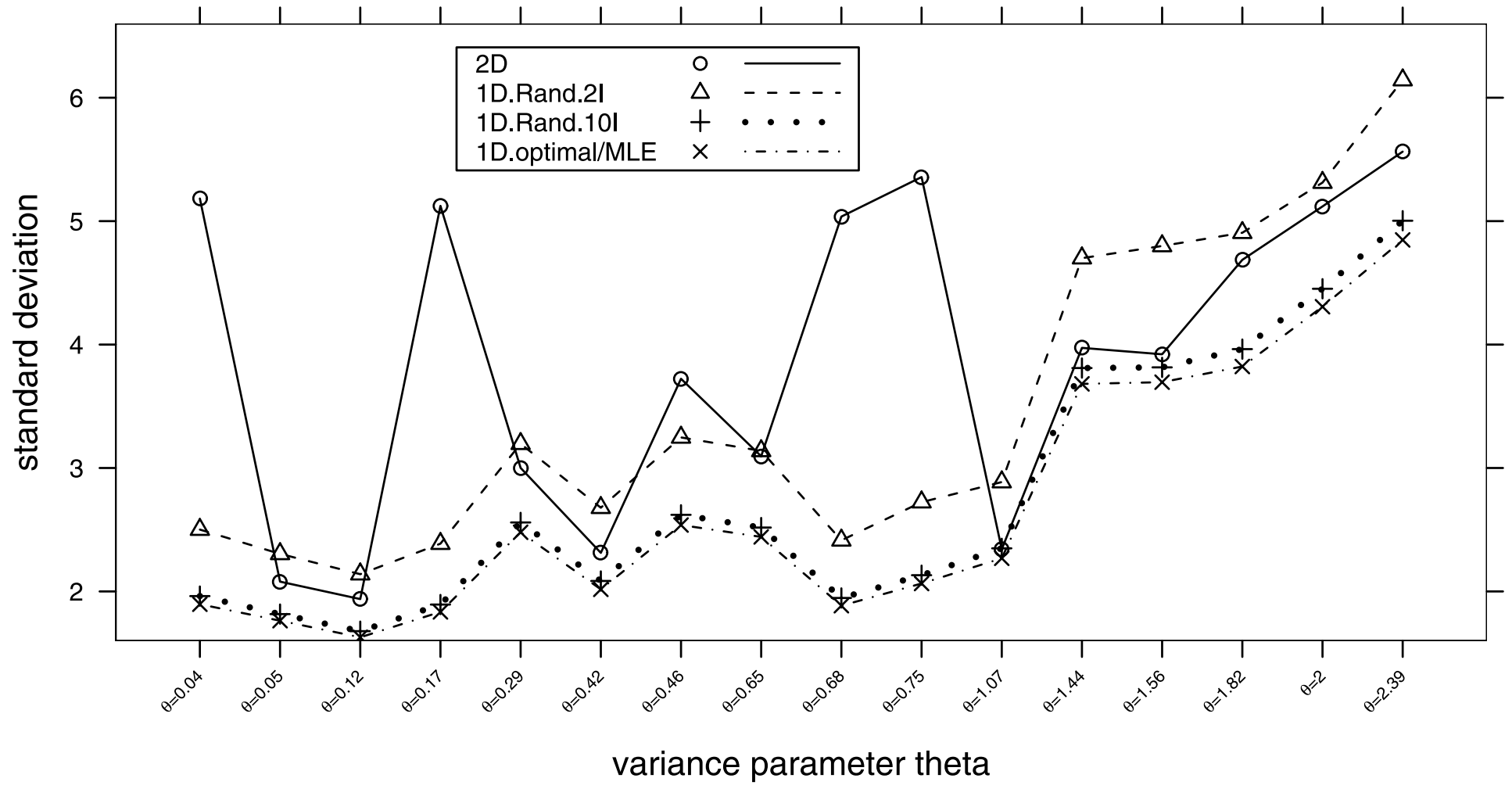

FIG 4. *The limit standard deviations of $\sqrt{n}(\hat{\theta} - \theta)$ for four estimators for the 16 variance parameters $\theta_i$: the two random 1D-projection estimators with $K = 2I = 32$ and $K = 10I = 160$ projections (reporting medians of 100 replications), $\hat{\theta}_{2D}$ and the optimal $\hat{\theta}_{1D}$. A is given in (13) below. The x-axis represents the values of 16 specified variance parameters, $\theta_i, i = 1, \ldots, 16$.*

where

$$\mathcal{C}_{2D}(a,b) = \frac{1}{2} \sum_{k<k'} \left( [A_{ka}, A_{k'a}][\Sigma_{kk'}^{kk'}]^{-1}[A_{kb}, A_{k'b}]^T \right)^2,$$

$$\mathcal{I}_{2D}(a,b) = \frac{1}{2} \sum_{k<k',l<l'} \left( [A_{ka}, A_{k'a}][\Sigma_{kk'}^{kk'}]^{-1}\Sigma_{kk'}^{ll'}[\Sigma_{ll'}^{ll'}]^{-1}[A_{lb}, A_{l'b}]^T \right)^2,$$

and $\Sigma_{kk'}^{ll'} = \begin{bmatrix} \Sigma_{kl} & \Sigma_{kl'} \\ \Sigma_{k'l} & \Sigma_{k'l'} \end{bmatrix}$ is a $2 \times 2$ matrix consisting of the elements of $\Sigma$.

Since it is hard to evaluate in closed form the expectation of the asymptotic covariance matrix of $\hat{\theta}_{1D}$ using random 1D projections, we use simulations to study its performance. In the simulation, we use a $7 \times 16$ $A$ matrix, shown in (13) below, derived from a later simulation study of a traffic demand tomography problem on a one-router network in Figure 3. That is, $I = 16$ and $J = 7$. The parameters $\theta = (\theta_1, \ldots, \theta_{16})^T$ are generated i.i.d. from the exponential distribution with mean 1, and remain fixed throughout the simulation. In each simulation run, we randomly draw $K = 2I$ and $K = 10I$ 1D-projection directions $\beta_k$ as in (10), and then calculate the limit covariance matrix of these two estimators using (5). We then replicate this procedure 100 times. For comparison, we also calculate the limit covariance matrix of $\sqrt{n}(\hat{\theta}_{1D} - \theta)$ for the optimal 1D-projections using (9) (same as that of $\sqrt{n}(\hat{\theta}_{MLE} - \theta)$), and that of $\sqrt{n}(\hat{\theta}_{2D} - \theta)$. Figure 4 shows the limit standard deviations of the 16 variance parameters $\theta_i$ for four estimators: the two random 1D-projection estimators (reporting medians of 100 replications), $\hat{\theta}_{2D}$ and the optimal $\hat{\theta}_{1D}$. The plot shows that the 2D-projection estimator is not efficient asymptotically by comparing with the optimal one or MLE. The plot also suggests that as the number of projections grows, the limiting covariance matrix for the random 1D-projections converges to the optimal one but we leave it to future studies.

## 4. Simulation studies for realistic models in network tomography

In Section 3, we have shown that for the simple Gaussian tomography model (4), the estimator $\hat{\theta}_{1D}$ by using the optimal set of 1D-projections of $\mathbf{Y}$ is asymptotically as efficient as MLE (Theorem 3.1). For other models, however, it is not clear whether the 1D-projection method can lead to an efficient estimator since a linear structure inherent in the Gaussian model no longer exists. In addition, the correlation rule which gives an optimal set of 1D-projections (9) may no longer be optimal. Since theoretical investigations of efficiency are difficult for general models, we resort to simulations to study the performance of the 1D-projection method (i.e. $\hat{\theta}_{1D}$ and $\hat{\theta}_{CF}$), especially using 1D-projections that obtained from the correlation rule (9). We study the performance of the method under two realistic models in network tomography. The first example is for traffic demand tomography [5, 21] with Gaussian OD traffic model where the mean and variance of traffic counts is related by a power equation [5], and the second example is for delay tomography [16, 18, 20] with mixture models for link delay distributions. We demonstrate that in both cases the 1D-projection method yields satisfactory results.

### *4.1. Traffic demand tomography using the Gaussian OD traffic model*

Let $\mathbf{X} = (X_1, \ldots, X_I)^T$ be the Origin-Destination (OD) traffic counts in a network. In traffic demand tomography, we observe $\mathbf{Y} = A\mathbf{X}$, where $\mathbf{Y}$ is the measured link traffic counts collected for instance using SNMP, and $A$ is the network routing matrix with elements 0 or 1. The problem is to estimate the distribution of each $X_i$ from independent samples of $\mathbf{Y}$. Following [5], assume a Gaussian OD traffic model with a power relation between the variance and mean of traffic counts, i.e., each $X_i$ $(i = 1, \ldots, I)$ is an independent Gaussian random variable with $var(X_i) = \phi(EX_i)^c$, where $\phi$ is an unknown scale parameter and $c > 0$ is a known power exponent. Let $\theta_i = EX_i$ be the mean counts of the $i$th OD traffic. The parameter of interest is $\theta = (\theta_1, \ldots, \theta_I)^T$ and $\phi$ is an unknown nuisance parameter.

In the simulation, we use the same one-router network with four attached input/output links as in [5], reproduced here in Figure 3. For this network, there are a total of 16 OD pairs from all pairwise combinations of an input and output link. For a certain arrangement of the 16 OD pairs, as described in [5], the routing matrix $A$ can be written as a $7 \times 16$ matrix that has entries of 0 except in the places indicated below

$$(13)\qquad A = \begin{bmatrix} 1&1&1&1&&&&&&&&&&&& \\ &&&&1&1&1&1&&&&&&&& \\ &&&&&&&&1&1&1&1&&&& \\ &&&&&&&&&&&&1&1&1&1 \\ 1&&&&1&&&&1&&&&1&&& \\ &1&&&&1&&&&1&&&&1&& \\ &&1&&&&1&&&&1&&&&1& \end{bmatrix}.$$

For each OD pair $X_i$, we generate the mean traffic rate $\theta_i$ independently from a lognormal distribution with mean=10 and sd=.95 in the log scale. The variance of the traffic rate is generated from a mean-variance relation $var(X_i) = 1000\theta_i$, i.e., the scale parameter $\phi = 1000$ and power exponent $c = 1$. These parameters are chosen to be compatible with the real data observed on the same one-router network in [5].

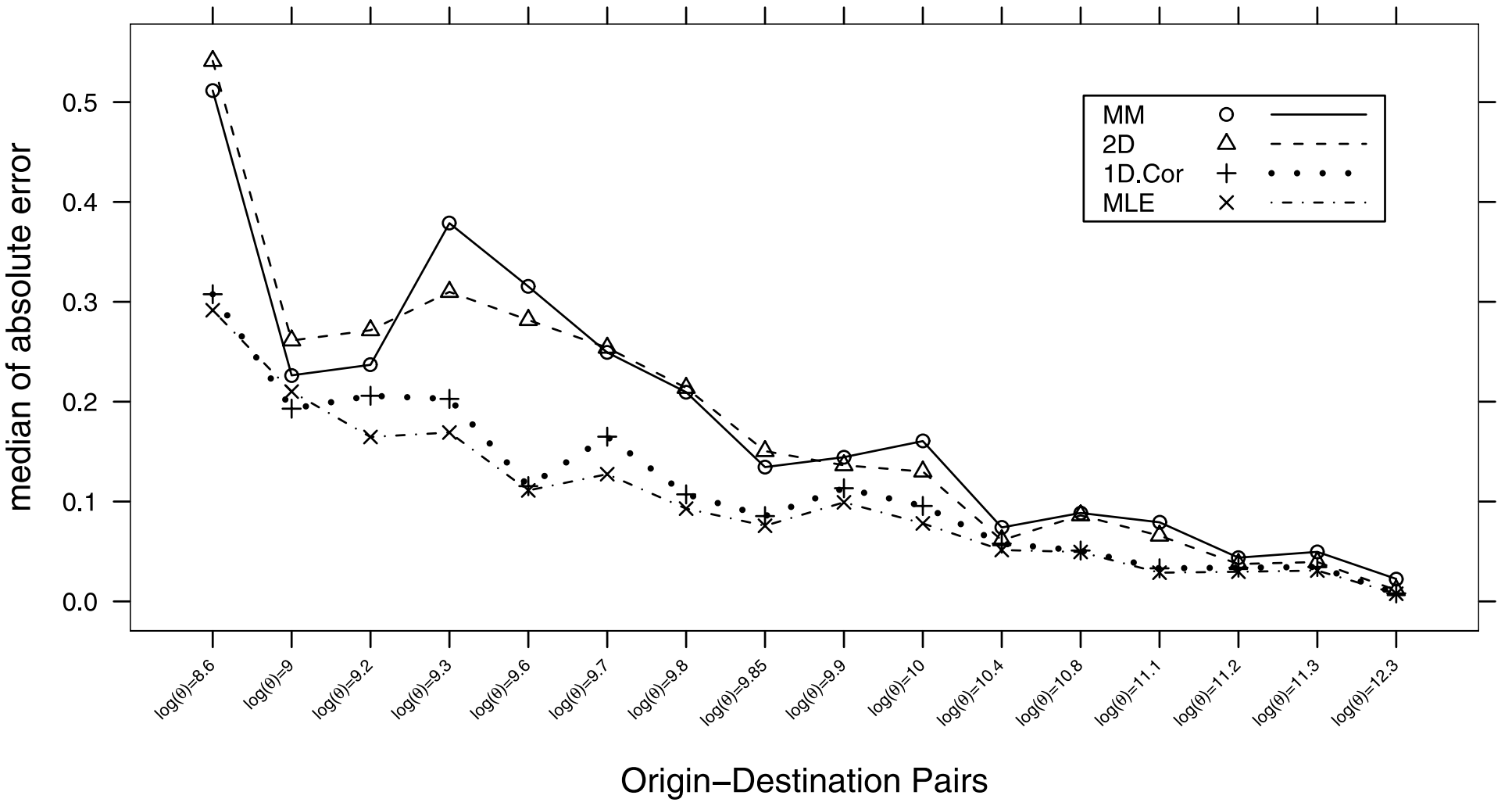

FIG 5. *The median estimation errors of mean traffic rates for the 16 OD pairs from 100 simulation runs. The estimation error is measured by* $|\log \hat{\theta}_i - \log \theta_i|$, $i = 1, \dots, 16$, *The four estimates are: moment estimator (MM), estimates obtained from the marginal likelihoods of 2D-projections (2D), from the marginal likelihoods of 1D-projection method using the optimal set of projections in (9) (1D), and MLE. The x-axis represents the specified mean traffic intensity values for the 16 OD pairs in a log based scale. Details of the simulation setup are described in Section 4.1.*

In each simulation run, we generate $n = 1000$ independent samples of OD traffic counts and estimated the mean OD traffic rates from the resulting link traffic counts. We consider four estimators for performance comparison: MLE, likelihood based estimator obtained from the correlation-based 1D-projections, likelihood based estimator obtained from all pairwise 2D-projections as proposed in [16] and a moment estimator. Similar to that used in [21], the moment estimator here is obtained by solving a system of over-determined linear equations constructed using the mean and variance of the link traffic counts $\mathbf{Y}$. The moment estimator is also used as the starting value of the other estimators obtained using numerical optimizations.

Figure 5 shows the median estimation errors of mean traffic rates for 16 OD pairs from 100 simulation runs. The estimation error for each $\theta_i$ is measured by $|\log \hat{\theta}_i - \log \theta_i|$ for all estimates. The plot shows that the correlation based 1D-projection estimator performs slightly worse than MLE, but much better than the 2D-projection estimator and the moment method. We have observed very similar results for other parameter settings. This suggests that the correlation based 1D-projections may be close to being optimal for the Gaussian OD traffic model with power mean-variance relation.

### *4.2. Network delay tomography using mixture modeling of link delays*

In the context of network delay tomography, one is interested in inferring the distributions of network link delays[3] from the end-to-end delay measurements, obtained either passively or actively. As an example, consider the four-leaf network shown in Figure 2. Here $I = 7$ and $J = 4$. Suppose that at the root of the tree, we send active multicast probes to the four leaves, resulting four simultaneous end-to-end delay measurements for each probe. Let $\mathbf{X} = (X_1, \ldots, X_7)^T$ be the internal link delays, and $\mathbf{Y} = (Y_1, \ldots, Y_4)^T$ be the end-to-end delay measurements. Then $\mathbf{Y} = A\mathbf{X}$, where $A$ is the routing matrix with elements 0 or 1 derived from the tree. The network delay tomography problem is to infer the distribution of $\mathbf{X}$ from independent samples of $\mathbf{Y}$.

[3]To be precise, the delay here is the queuing delay that excludes the constant link propagation delay, we omit queuing when context is clear.

There have been significant amount of work on network delay tomography [3, 10, 16, 18, 20], and most of these approaches use likelihood based estimation. The likelihood based estimation uses a discrete distribution with equal width bins to approximate a continuous link delay distribution, where the same bin width is used for all the links. However, for a heterogeneous network such as the Internet, such a discretization process will result in an over-parameterized link delay model and hence lead to high computational complexity. Recently, a characteristic function based estimation approach has been proposed as an alternative approach to accommodate network heterogeneity ([9]), using a flexible mixture modeling of link delays. As described in Section 2.2.2, instead of minimizing the Kullback-Leibler distance between the empirical and model distribution, by which the maximum likelihood estimator is derived, their estimator is derived by minimizing a $L_2$ distance between empirical and the model characteristic function. It is also found that the estimator derived from comparing the characteristic functions of low dimensional components of $\mathbf{Y}$ yields better performance, as compared to $\mathbf{Y}$ itself, where the difficulty of the latter is how to choose an appropriate *high dimensional* weight function $\mu$.

In the following, we run simulations of the delay tomography model on the four-leaf tree, and compare the performance of $\hat{\theta}_{CF}$ defined in (3) based on 1D-projections with that based on all pairwise 2D-projections of $\mathbf{Y}$. We do not compare the estimates against MLE because MLE is computationally infeasible for the flex-ible mixture model that we use in deriving these estimates. In the simulation, each link delay distribution $X_i, i = 1, \ldots, I$, is generated independently from a queuing distribution of an M/M/1 queue of the following form [11]

$$P(X_i > x) = u_i \exp(-v_i^{-1} x), \quad x > 0, \quad \text{and} \quad P(X_i = 0) = 1 - u_i$$

where $0 < u_i < 1$ is the utilization of the queue, and $v_i^{-1} > 0$ is the service rate of the queue time $(1 - u_i)$. For each link delay $X_i$, we generate a corresponding $u_i$ from a uniform distribution in the interval [0.3,0.7], and $v_i$ from an exponential distribution with mean 3. To obtain our estimates, we first model each link delay by a piecewise uniform distribution with an exponential tail, using 10 bins placed at quantiles of the distribution[4]. The estimates of the link delay distributions based on 1D-projections of $\mathbf{Y}$ are obtained by (3), where a Gaussian weight function, with standard deviation of 5 after normalizing each projection, is used for $\mu$ to calculate the $L_2$ distance and the integrals are approximated by using Monte-Carlo methods. The estimates based on 2D-projections are computed similarly.

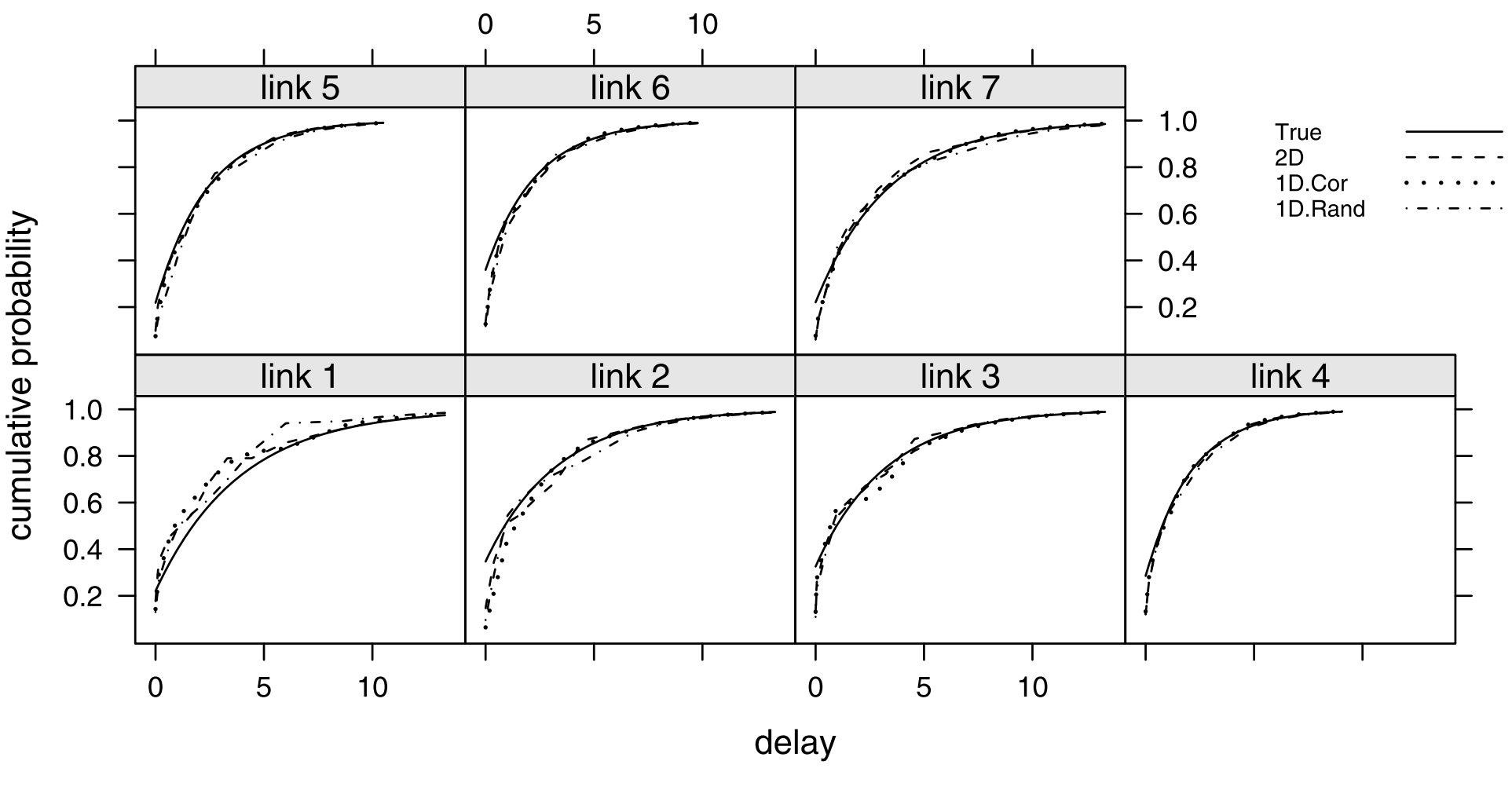

Fig 6. *The estimated cumulative distribution functions of link delays on a four-leaf tree (Figure 2) from 1000 end-to-end delay measurements. The true delay distributions, shown in solid line, are those generated from a M/M/1 queue. All three estimates are obtained using the marginal characteristic function based approach described in Section 2.2.2 using a mixture model of piecewise uniform of 6 bins and an exponential tail. The estimates are: 2D-projection method (2D), 1D-projection method using projections in (9) (1D.Cor), and 1D-projection method using I random projections in (9) (1D.Rand). Details of the simulation setup are described in Section 4.2.*

In each simulation run, a total of 1000 delay samples are generated for each link in the four-leaf tree from its specified delay distribution and then the end-to-end delays are computed. For each of the seven link delay distributions, we consider three estimators: $\hat{\theta}_{CF}$ by using the correlation based 1D-projections as in (9), $\hat{\theta}_{CF}$ using $K = I = 7$ random 1D-projections adjusted for the covariance of $\mathbf{Y}$, and the characteristic function based estimator using all pairwise 2D-projections by [16]. Figure 6 plots the cumulative distributions of the estimated link delays along with the generated true distribution for one simulation run. From the figure, we observe that all estimators yield satisfactory results.

To measure the accuracy of the estimates, we use the Mallows' distance [17] defined for a cumulative distribution distribution $F$ and its estimate $\hat{F}$ by

$$M(F, \hat{F}) = \int_0^1 \left| F^{-1}(p) - \hat{F}^{-1}(p) \right| dp,$$

where $F^{-1}$ and $\hat{F}^{-1}$ are the inverse cumulative distributions. The Mallows' distance can be viewed as the average of absolute difference in quantiles between two distributions. Because the Mallows' distance is linear to the scale of distributions, we use $M(F, \hat{F})/\sigma_F$, the normalized Mallows' distance, to measure the difference between $F$ and the estimate $\hat{F}$, where $\sigma_F$ is the standard deviation of $F$.

We repeat the simulation 100 times and compute the normalized Mallows distance between the estimated and the simulated distributions for all links. Figure 7 reports the median of the normalized Mallows' distance between the three estimates and the generated true distributions for each link. We observe that the estimator using 2D-projections yield best results overall, the estimator using the correlation based 1D-projections is a close second, and that using random 1D-projections performs the worst. This indicates that although the correlation based 1D-projections may not be the optimal directions, the information loss using these 1D-projections as compared to all pairwise 2D-projections is not significant.

[4]The quantiles are unknown in real applications, but they are assumed to be predetermined here for simplicity. Otherwise, they can be estimated through iterations of the estimation process with an initial value. This has been suggested in [9] as a strategy for placing bins for modeling the link delays and has been shown to yield accurate estimators.

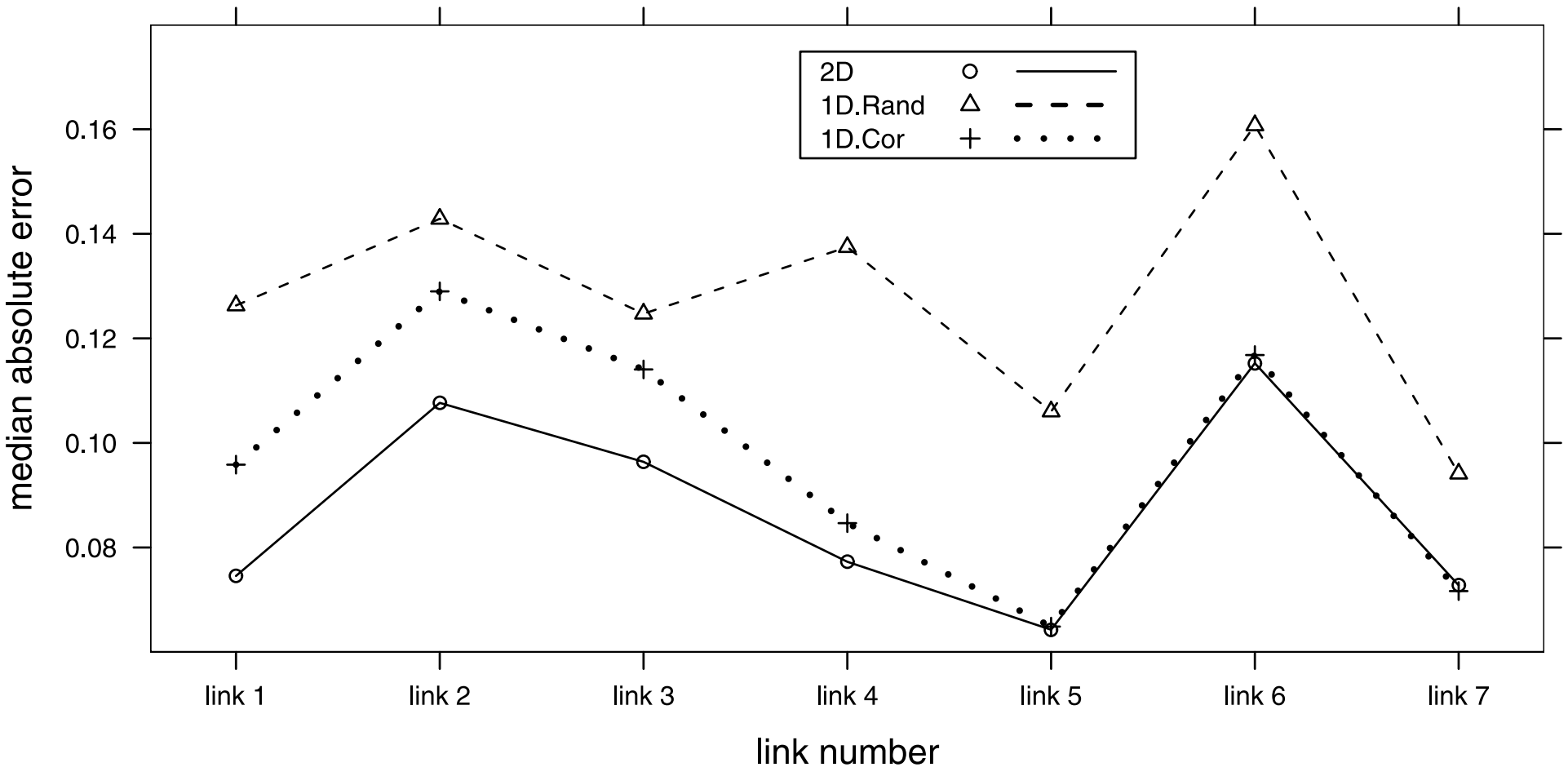

Fig 7. *Medians of the normalized Mallows' distance for the three link delay distribution estimates from 100 simulation runs, where each simulation run the same setup as in Figure 6.*

## 5. Conclusion and future work

This paper proposes a one-dimensional linear projection method for solving a class of linear inverse problems in network tomography. For the simple Gaussian tomography model, the optimal set of 1D-projections is derived and it is shown that a likelihood based estimator based on these 1D projections is asymptotically as efficient as the usual maximum likelihood method. For more realistic models in network tomography, simulation studies show that the estimators derived from the marginal distributions or marginal characteristic functions of 1D-projections perform well for large sample sizes. Future work includes generalization of the optimal design of 1D-projections for non-Gaussian tomography models and small sample studies of the proposed method.

**Acknowledgments.** We would like to thank Gang Liang at UC Irvine for sharing his code of their pseudo likelihood method and to David James in our department for helpful comments on the paper.